\def\year{2022}\relax
\documentclass[letterpaper]{article} 
\usepackage{aaai22}  
\usepackage{times}  
\usepackage{helvet}  
\usepackage{courier}  
\usepackage[hyphens]{url}  
\usepackage{graphicx} 
\usepackage{natbib}  
\usepackage{caption} 
\DeclareCaptionStyle{ruled}{labelfont=normalfont,labelsep=colon,strut=off} 
\usepackage{algorithm}
\usepackage{algorithmic}
\usepackage{booktabs}       
\usepackage{float}       
\usepackage{fancyhdr}
\usepackage{caption} 
\usepackage{subcaption} 

\usepackage{newfloat}
\usepackage{listings}
\floatstyle{ruled}
\newfloat{listing}{tb}{lst}{}
\floatname{listing}{Listing}

\newcommand\blfootnote[1]{%
  \begingroup
  \renewcommand\thefootnote{}\footnote{#1}%
  \addtocounter{footnote}{-1}%
  \endgroup
}

\title{FingerFlex: Inferring Finger Trajectories from ECoG signals}
\author {
    Vladislav Lomtev,\textsuperscript{\rm 1}\textsuperscript{\rm 2}
    Alexander Kovalev, \textsuperscript{\rm 2}
    Alexey Timchenko \textsuperscript{\rm 3}\textsuperscript{\rm 4}
}
\affiliations {
    \textsuperscript{\rm 1} Bauman Moscow State Technical University, Moscow, Russia\\
    \textsuperscript{\rm 2} ALVI Labs \\
    \textsuperscript{\rm 3} Brain Dynamics Group, Higher School of Economics, Moscow, Russia \\
    \textsuperscript{\rm 4} University of Tuebingen, Tuebingen, Germany \\
    vladlomtev@gmail.com, koval.alvi@gmail.com, aleksejs.timcenko@gmail.com
}

\usepackage{tabularx} 
\usepackage{dblfloatfix}

\begin{document}
\maketitle

\begin{abstract}
Motor brain-computer interface (BCI) development relies critically on neural time series decoding algorithms. Recent advances in deep learning architectures allow for automatic feature selection to approximate higher-order dependencies in data. This article presents the FingerFlex model - a convolutional encoder-decoder architecture adapted for finger movement regression on electrocorticographic (ECoG) brain data. State-of-the-art performance was achieved on a publicly available BCI competition IV dataset 4 with a correlation coefficient between true and predicted trajectories up to 0.74. The presented method provides the opportunity for developing fully-functional high-precision cortical motor brain-computer interfaces.

\end{abstract}
\blfootnote{Preprint. Under review}

\section{Introduction}
\subsection*{Brain-computer interface}

One of the promising rehabilitation tools for people with disabilities is the brain-computer interface (BCI) \citep{lebedev2006, bamdad2015, yuan2014}. In general, developing a functioning BCI involves interpreting neural signals to extract relevant information from the brain. The extracted features, i.e. control parameters can be used in the restoration of the lost functions and ability augmentation via controlling external devices. Specifically, motor BCIs aim to decode movement-related information in the sensory-motor cortex into meaningful behavioral patterns, such as joint movement or finger flexion \citep{tam2019, volkova2019}. 

\subsection*{Recording of the neural activity}

Brain activity is analyzed using electrophysiological methods measuring single neuron or population level electric potentials. Most methods allowing to capture spiking activity, such as optogenetics, single-unit and multi-unit recordings are highly-invasive. At the same time, population level recording tools, such as electroencephalogram (EEG) and magnetoencephalogram (MEG) require no surgical intervention, albeit lack spacial resolution due to a voltage leakage in the human scalp and movement-related artifacts \citep{neilcuffin1979, hamalainen1993, jackson2014}. The electrocorticogram (ECoG) used in this study lies in between these methods, estimated to register localized activity of ~$10^5$ neurons on the cortical surface \citep{engel2005}. ECoG electrodes are placed over the surface of the cortex with approximately $1$ cm inter-electrode distance. Electrocorticogram provides higher spatial resolution and superior signal quality compared to EEG and MEG with a tradeoff of portability and invasiveness.

\subsection*{Movement correlates in the human brain}

The primary brain area involved in the movement processing is primary motor cortex (PMC) \citep{kakei1999}. The adjacent brain areas -  supplementary motor area (SMA) and somatosensory cortex take part into motor planning and receive peripheral inputs on the body position and environment \citep{tam2019}. Planning and executing motor programs change the activity of single neurons in the cortical sheet of the motor areas \citep{kakei1999}. Spiking activity in a synchronized neuronal population, in turn, contributes to the change in local field potentials registered via surface electrodes \citep{buzsaki2012}. Despite the lack of an exact area-to-muscle correspondence in the motor cortex, overlapping regions associated with different muscle groups and even fingers can be distinguished \citep{schieber2001,marjaninejad2017,georgopoulos2015}.

Another feature of population activity in the cortex is that its local field potential comprises a set of independent oscillatory processes \citep{buzsaki2004}. It was shown that changes in alpha band (8-13 Hz) power, also known as mu-rhythm, beta (13-35 Hz), gamma (35-100 Hz) and high-gamma (100-250 Hz) correlate significantly with ongoing movement patterns \citep{tam2019, marjaninejad2017}. It is thus possible to use frequency-decomposed or time-frequency represented signal instead of raw time series data. 

The primary purpose of this study is to improve existing solutions for decoding a continuous finger trajectory based solely on ECoG-registered brain oscillatory patterns in the sensorimotor cortex. 

\subsection*{Related work}

While some studies concentrated on classifying finger activations \citep{hotson2016, onaran2011, yao2019}, the others predicted hand translations \citep{pistohl2008, nakanishi2013, bundy2016, sliwowski2022} and gestures \citep{pan2018, branco2017}, numerous approaches were presented for solving finger trajectory prediction task. 

\citet{liang2012} and \citet{flamary2012} presented the first attempts to decode one-dimensional finger flexions on the publicly available BCI Competition dataset IV \citep{schalk2007}. The former study employed switching linear models, while the latter - linear regression with rigorous feature selection, which allowed to win the competition. \citet{nakanishi2014} used sparsed linear regression on frequency band decomposed signal achieving comparable results both on the BCI competition dataset and data gathered by their team. With the advancement of deep learning \citep{lecun2015} techniques over the following years, neural network approaches were more frequently used in consequent studies. For example, \citet{xie2018} used long short-term memory based \citep{hochreiter1997} architecture to decode finger trajectory. \citet{jubien2019} compared different conventional machine learning methods with artificial neural network approaches on the same task to conclude that deep learning solves the task with higher accuracy. The method developed by \citet{frey2021} presented the opportunity for developing a generalized decoder for neural data yielding high performance over a variety of tasks including finger trajectory regression over the ECoG data. The attempt to build an interpretable convolutional neural network was made by \citet{petrosyan2021}, which is highly relevant for neuroscience research on details of motor encoding in the human brain. Lastly, a new type of features based on Riemann geometry \citep{congedo2017} was introduced in the finger movement decoding task by \citet{yao2022} used consequently with linear discriminators and support vector machine (SVM) regressors. 

The novelty of this work is primarily addressed by the use of different decoding paradigm based on the encoder-decoder architecture, which transforms feature time series into a target time series through a set of convolutions, deconvolutions, and skip-connections. Our research aims to prove this approach can outperform existing conventional machine learning solutions based on thorough feature extraction, as well as deep learning models implemented by the other research teams in a finger flexion decoding task.

\section{Methods}

\subsection{Dataset}

The primary dataset for training the model was the BCI competition IV dataset 4 \citep{schalk2007}. The dataset contains electrocorticographic recordings of three anonymous patients with corresponding finger movement information, represented as a one-dimensional coordinate denoting the finger fold. The recording grids consist of 48-64 electrodes evenly distributed in grids of 6x8, 8x8, and other shapes. The authors of a dataset imposed a limitation on ECoG electrode distribution information. The channels were shuffled beforehand and the electrode positions relative to the brain were not provided either. ECoG was recorded at a sampling frequency of 1000 Hertz, while finger movements were sampled at 25 Hz with a data glove.

The subjects were instructed to move a particular finger after a cue, which was presented as a corresponding word (e.g., ”thumb”) on a computer monitor. Each cue lasted for two seconds and was followed by a two-second resting period during which the screen was blank. More details on the experimental procedure and recording setup can be found in the dataset description \citep{miller2008}.

\subsection{Data preprocessing and feature extraction}

\begin{figure}[ht]
    \centering
    \includegraphics[width=0.9\columnwidth]{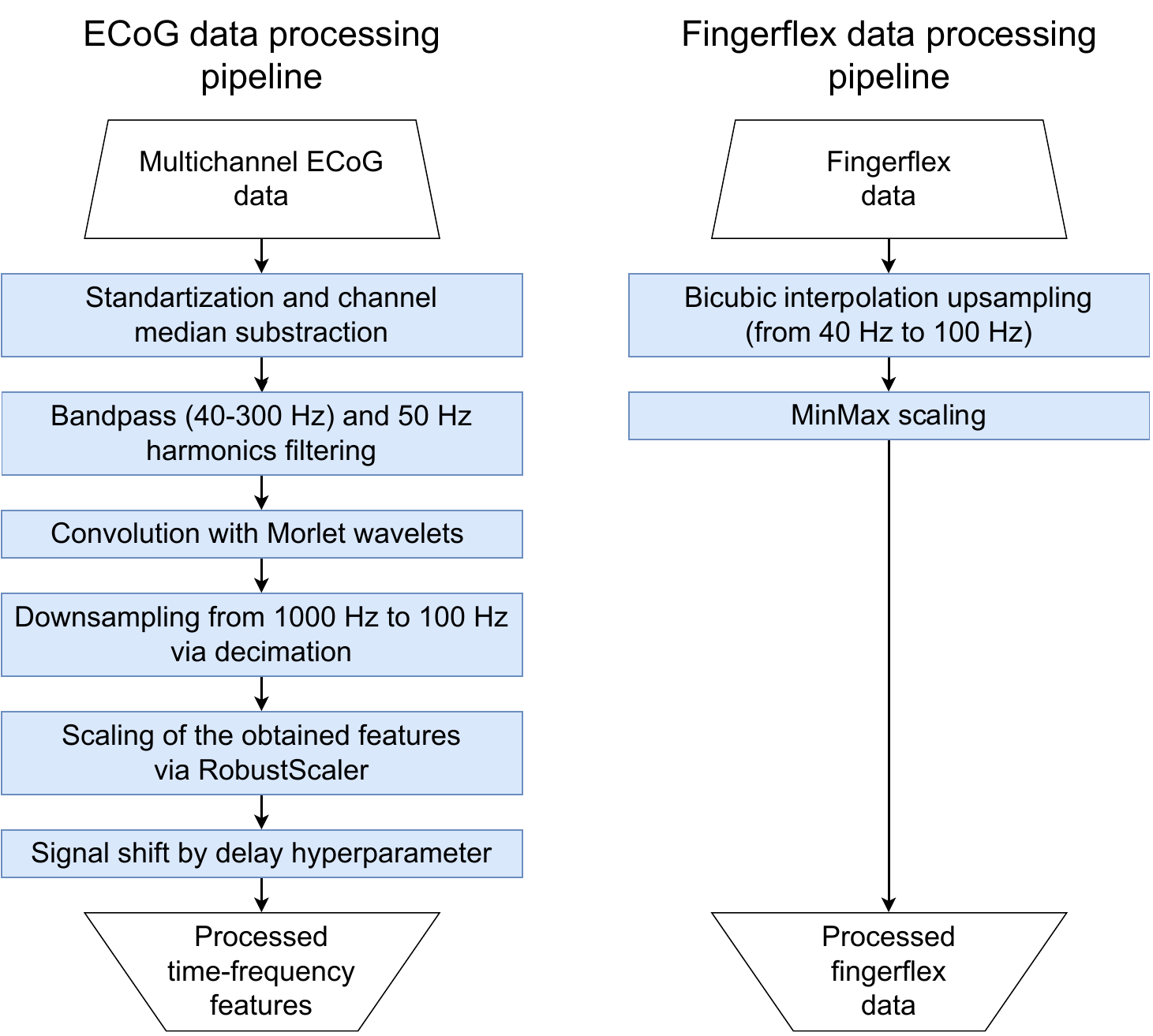}
    \caption{ECoG and finger motion data processing pipeline}
    \label{fig:processing_pipeline}
\end{figure}

Data preprocessing can be divided into two parts: preprocessing of ECoG signals and preprocessing of the finger movement data (Figure \ref{fig:processing_pipeline}). The ECoG preprocessing pipeline consisted of the following steps. Firstly, data standardization and subtraction of the median from each channel was performed. Then, band-pass and band-stop filters were used to filter frequencies in the range of 40 to 300 Hz and remove the power grid frequency of 50 Hz and its harmonics. The frequency range from 40 to 300 Hz constitutes gamma and high-gamma components of the local field potential signal, which were chosen as the main predictors for the task.

As a method of explicit extraction of time-frequency representation of the signal, the convolution with complex Morlet wavelets was chosen. Wavelet decomposition allows capturing specific frequency components of a signal change in time. Forty wavelet kernels were created with base frequencies ranging from 40 to 300 Hz evenly spaced on a logarithmic scale. The convolution operation of selected wavelets with each channel was performed and the amplitudes of resulting analytical signals were extracted as an absolute value of a complex-valued vector. As a result of this operation for each ECoG channel, a spectrogram is obtained of shape: (number of wavelets, number of time points in a time window). Then, the sampling rate was then lowered from 1000 Hz to 100 Hz by decimation for the correspondence to a motion capture time series. Next, the obtained time-frequency envelopes were standardized with respect to $0.1$ and $0.9$ quantiles, and the values not falling within this range were then taken equal to the corresponding scaler boundary values to combat outliers. This operation was done using the RobustScaler class from the scikit-learn library, which was fitted on the training set and applied to both the training and validation sets. Lastly, to account for the delay between brain activity and actual movements, the forward shift of ECoG data with respect to the finger movement time series was introduced. The exact value of the shift can be regarded as a tunable hyperparameter within the relevant physiological range: $0 - 200$ms.

The preprocessing of finger movement data consisted of two steps. First, to keep the sampling frequencies of finger movements and time-frequency time series matched,  upsampling from $25$ Hz to $100$Hz using bicubic interpolation was conducted. Second, MinMax scaling was applied.

\subsection{The model architecture}

\begin{figure*}[ht]
    \centering
    \includegraphics[width=0.9\linewidth]{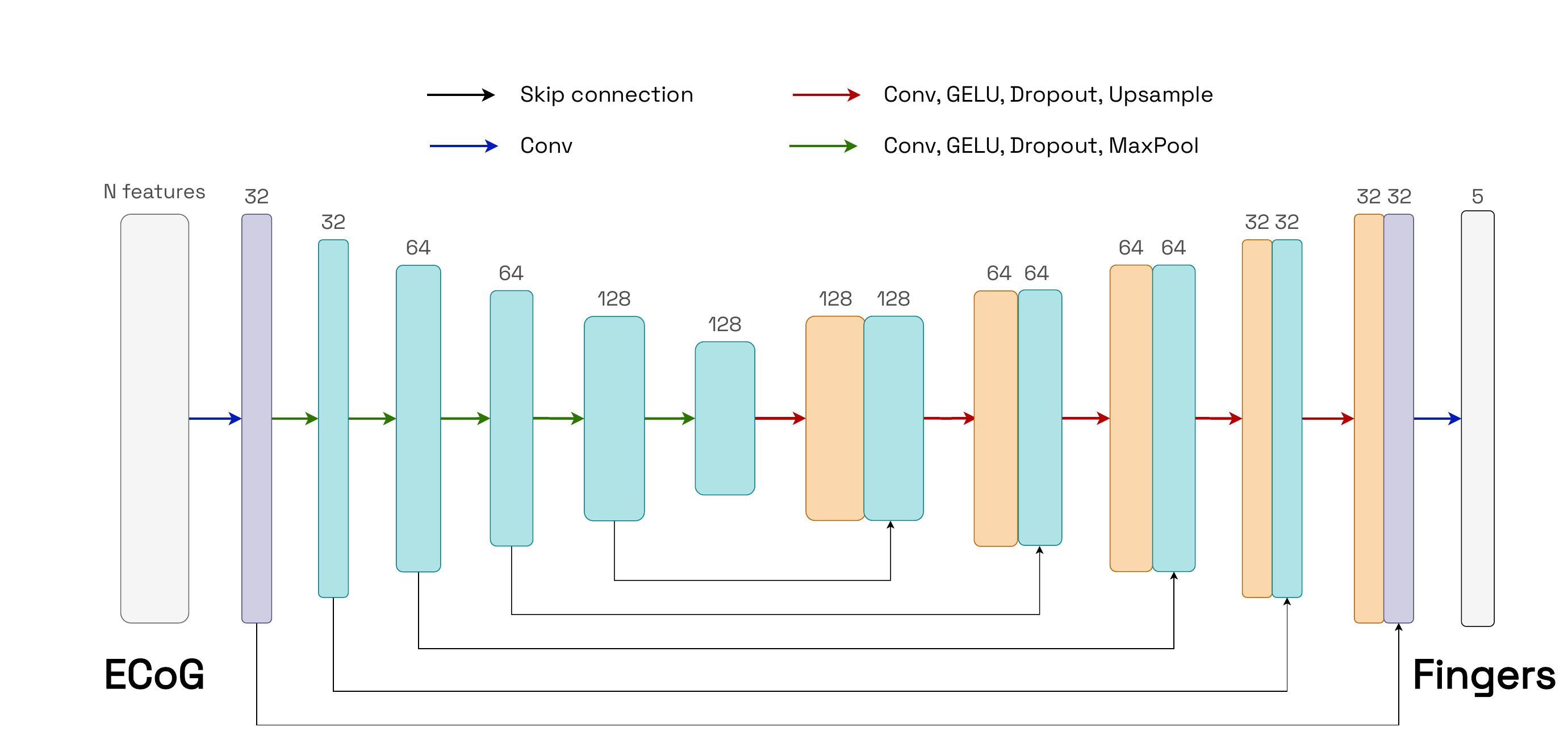}
    \caption{FingerFlex model architecture}
    \label{fig:model_architecture}
\end{figure*}

The final model is designed in the same way as the convolutional autoencoder with a set of additions and improvements. The architecture of the model is depicted on Figure \ref{fig:model_architecture}

As the input, the model takes a window of ECoG features of arbitrary length and returns the corresponding window of predicted movements; however, during the training process, the length of such a window is fixed and equal to $256$ time points ($2.56$ secs). Initially, the model performs dimensionality reduction using a regular convolutional layer. Next comes a set of convolutional encoder layers. Each encoder layer consists of a 1D convolution layer, layer normalization, GeLU activation function, dropout layer, and maximum pooling with stride equals $2$. This part of the model is responsible for encoding the features of the time window fragments. Encoder layer architecture is inspired by wav2vec architecture \citep{schneider2019}.

This is followed by a symmetrical set of convolutional decoder layers, each of which uses for further prediction not only information from the previous layer, but also from the symmetrical encoder layer. In other words, skip connections like those of the UNet model \citep{ronneberger2015} were added. Skip connections make it possible to use for prediction not only the generalized representation in the latent space obtained by the encoder, but also the original local features, which ultimately increases the accuracy of the final model. The model is completed by a $1\times1$ convolution layer that uses the resulting set of features from each point in time to predict the coordinates of each of the five fingers at that point.

\subsection{Training details}

The BCI competition IV dataset 4 was already divided into training and validation sets, the recording of each of the three patients was divided into $2$ parts, where the first part of $6.5$ minutes duration constitutes training subset and the second part of $3.5$ minutes long constitutes validation data subset. The model was trained and optimized using only the training subset, while the validation set was used for the performance measurement and hyperparameter adjustment. Due to lack of alignment of electrode grid shapes between subjects and intentional mixing of electrodes done by the dataset creators, the model was trained for each subject individually. Note that overfitting due to a hyperparameter search using the validation set was addressed by performing ablation study using only one subject provided in the dataset. The same model was trained on the other patients without any additional hyperparameter adjustment.

\paragraph{Loss function}

The loss function was chosen as the half sum of the mean squared error (MSE) and the mean cosine distance. Since the metric of the final score is correlation, it is essential to include it in the loss for training process. However, the correlation itself is not an optimal optimization criterion when working with small windows, since it is too sensitive to insignificant noise within this window in those cases when there is no movenet. Therefore, instead of a correlation loss, it was decided to use the cosine distance. The ablation study was conducted to test whether the combination of MSE and cosine similarity metrics performed better than any of those metrics alone. 
Nevertheless, due to the evaluation criteria at BCI competition IV, correlation coefficient (CC) metric was used to evaluate and compare model performance.

\paragraph{Hyperparameters}

The number of features on each of the layers of the encoder as a result of the experiments was taken equal to $(32, 32, 64, 64, 128, 128)$. The number of features in the decoder layers is symmetrical with respect to the bottleneck. The stride size was taken equal to $2$ to reduce the sampling rate of the original window as gradually as possible, which allows better extraction of common features.

The learning rate was picked according to the Pytorch Lightning library's automatic learning rate finder recommendation and was set fixed to $8.4 \cdot 10^{-5}$.

As was previously mentioned, wavelet number, loss function combination coefficients and the delay between ECoG and hand movement time series can be considered as the tunable hyperparameters.

\section{Results}

\subsection{Model development}
Every subsequently improved model was trained for $30$ epochs. Model weights trained during the epoch preceding overfitting of the model were chosen for inference. The model refinement took several important steps to achieve the final performance.

\subsubsection{Architecture refinement}
Baseline encoder-decoder architecture, trained on unscaled absolute values of wavelet-convoluted signals, yielded average correlation value of $0.2$ on the first subject (Table \ref{table:model_improvement}). However, adding data scaling both for ECoG features and finger movements improved results using the same architecture by more than two-fold, to $0.45$ on average on the first subject. It can be speculated that this win in performance is due to highly variable absolute values of the time-frequency representation of the signal. The electrophysiological recordings consistently show a $1/f$ relationship between signal spectral power and a corresponding frequency \citep{buzsaki2004}, leading to a multiple-fold difference between signal power in highly distinct frequency bands. A modification of the encoder-decoder architecture parameters, which are mainly a number of features in each layer and the number of layers allowed to increase the average correlation coefficient further to $0.55$. The final model architecture is shown in Figure \ref{fig:model_architecture}. Then the use of wavelets only from the gamma frequency band and an increase in their number up to $40$ also led to an improvement in the results to an average correlation value of $0.6$. Finally, skip connections were added, which increased the mean correlation coefficient to $0.66$. 

\begin{table}[!h]
  \caption{\label{table:model_improvement}
  Model development steps with corresponding performance obtained on the validation set.} 
  \label{table:model_improvement}
  \centering
  \begin{tabular}{lllllll}
    \toprule
    Model improvement step & Mean correlation  \\
    \midrule
    Baseline model & 0.2 \\
    + Feature scaling &0.45 \\
    + Architecture improvement &0.55 \\
    + Frequency range adjustment &0.59 \\
    + Optimizing number of wavelets &0.6 \\
    + Adding skip-connections &0.66 \\
    \bottomrule
  \end{tabular}
\end{table}

\subsubsection{Hyperparameter tuning}
Then an ablation study was performed to calculate the optimal value of the delay hyperparameter between the ECoG signal and the predicted motion. The baseline delay chosen as the initial one was $20$ ms, and as the consequent experiments showed that for the built model the value of delay is not crucial, as long as it is within $0-200$ms range. Therefore, the final shift value between ECoG and finger motion time series was kept at $20$ ms. 

Another ablation study was conducted to test whether the use of a combined loss function - the sum of cosine distance and MSE is empirically justified. Cosine distance, when used alone yielded a correlation value of $0.44$, which is a worse result than that of MSE ($0.64)$. However, using a combination of cosine distance and MSE increases the correlation by an average of $0.02$ compared to the regular MSE, summing up to 0.66. 

\subsection{Decoding performance}

Finger movement inference, carried out on three subjects presented in the BCI Competition IV dataset 4 showed the following results: $0.66$ average correlation on the first subject, $0.62$ on the second and $0.74$ on the third (see Table \ref{table:bcicomp_performance}). Recall that each patient had its own model trained, however, hyperparameter tuning was carried out only using the first subject. The mean correlation coefficient across all subjects is $0.67$. Correlation values for different fingers are in the range of $0.6-0.75$ for the first subject, $0.53-0.73$ for the second and $0.68-0.78$ for the third.

\begin{table*}[h!]
  \caption{ \label{table:bcicomp_comparison} 
  Performance comparison on works solving BCI Competition IV dataset 4 finger trajectory decoding task}
  \label{table:bcicomp_comparison}
  \centering
  \begin{tabular}{lllllll}
    \toprule
    \multicolumn{4}{r}{Mean CC by subject}\\                  
    \cmidrule(r){2-4}
    Method & S1 & S2 & S3 & Average\\
    \midrule
    \midrule
    Flamary (2012) - BCI Competition 2nd place
    & 0.48 & 0.24 & 0.56  & 0.43 \\
    Liang and Bougrain (2012) - BCI Competition winner
    & 0.45 & 0.39 & 0.59 & 0.48 \\
    Xie et al. (2018) - LSTM
    & 0.56 & 0.41 & 0.58 & 0.52 \\
    Frey et al. (2021) - Multi-purpose CNN
    & N/A & N/A & N/A & 0.52 \\
    Petrosyan et al. (2021) - Interpretable CNN 
    & 0.45 & 0.34 & 0.56 & 0.45 \\
    Yao et al. (2022) - lightGBM 
    & 0.52  & 0.47 & 0.61  & 0.53 \\
    
    \textbf{Our solution (FingerFlex)} 
    & \textbf{0.66} & \textbf{0.62} & \textbf{0.74} &\textbf{0.67} \\
    \bottomrule
  \end{tabular}
\end{table*}

\begin{table}[h]
  \caption{ \label{table:bcicomp_performance} FingerFlex model performance on the BCI competition IV dataset 4. Correlation coefficients are calculated separately for each subject and finger using the provided validation set. 
  }
  \label{table:bcicomp_performance}
  
  \begin{tabular}{ m{0.17\columnwidth} m{0.08\columnwidth} m{0.08\columnwidth}  m{0.08\columnwidth} m{0.08\columnwidth} m{0.08\columnwidth} m{0.08\columnwidth} }
    \toprule
    \multicolumn{7}{c}{Correlation coefficient}\\                  
    \cmidrule(r){2-6}
    Subject & Thumb & Index & Middle & Ring & Little & Avg. \\
    \midrule
    \midrule
    \textbf{S1} & 0.66 & 0.75 & 0.64 & 0.66 & 0.6 &\textbf{0.66} \\
    \textbf{S2} & 0.66 & 0.73 & 0.53 & 0.63 & 0.54 &\textbf{0.62} \\
    \textbf{S3} & 0.78 & 0.68 & 0.68 & 0.78 & 0.76 &\textbf{0.74} \\
    \textbf{Average} & \textbf{0.7} & \textbf{0.72} & \textbf{0.62 }& \textbf{0.69} & \textbf{0.63} &\textbf{0.67} \\
    \bottomrule
  \end{tabular}
\end{table}

\subsection{Model verification on a different dataset}
To check model stability across datasets, the built model was also tested on Stanford dataset \citep{miller2019}. Since the data representation in this dataset was the same as in BCI Competition IV dataset 4, the entire data processing pipeline (Figure \ref{fig:processing_pipeline}) remained unchanged. 

\begin{table}[h]
  \caption{ \label{table:stanford_performance} FingerFlex model performance (correlation coefficients) on the ECoG finger motion dataset provided by \citet{miller2019}.
  }
  \label{table:stanford_performance}
  \centering
  \begin{tabular}{m{0.17\columnwidth}                  m{0.08\columnwidth}                  m{0.08\columnwidth}                  m{0.08\columnwidth}                  m{0.08\columnwidth}                  m{0.08\columnwidth}                  m{0.08\columnwidth}}
    \toprule
    \multicolumn{7}{c}{Correlation coefficient}\\                  
    \cmidrule(r){2-7}
    Subject & Thumb & Index & Middle & Ring & Little & Avg. \\
    \midrule
    \midrule
    \textbf{bs} & 0.65 & 0.66 & 0.45 & 0.53 & 0.44 &\textbf{0.55} \\
    \textbf{cc} & 0.76 & 0.7 & 0.61 & 0.76 & 0.73 &\textbf{0.71} \\
    \textbf{zt} & 0.59 & 0.73 & 0.68 & 0.62 & 0.53 &\textbf{0.63} \\
    \textbf{jp} & 0.68 & 0.67 & 0.31 & 0.65 & 0.53 &\textbf{0.57} \\
    \textbf{ht} & 0.31 & 0.32 & 0.34 & 0.36 & 0.36 &\textbf{0.34} \\
    \textbf{mv} & 0.42 & 0.85 & 0.42 & 0.40 & 0.71 &\textbf{0.56} \\
    \textbf{wc} & 0.45 & 0.63 & 0.42 & 0.27 & 0.31 &\textbf{0.42} \\
    \textbf{wm} & 0.33 & 0.08 & 0.25 & 0.32 & 0.06 &\textbf{0.22} \\
    \textbf{jc} & 0.65 & 0.37 & 0.32 & 0.43 & 0.37 &\textbf{0.43} \\
    \bottomrule
  \end{tabular}
\end{table}

The feature processing, hyperparameters and model architecture were fixed. The obtained results are represented in the Table \ref{table:stanford_performance}. While there are subjects, which the model fails to recognize finger movements with correlations of $0.2 - 0.4$, there are patients, who have mean correlation coefficients between predicted and actual finger movements in the range of $0.5 - 0.7$. The low performance on some patients could be attributed to irrelevant positioning of the ECoG grid with regards to brain areas mostly involved during movement planning and execution.

    

\section{Discussion}

\subsection{Contribution}

\begin{figure*}[h]
    \centering    \includegraphics[width=0.75\linewidth]{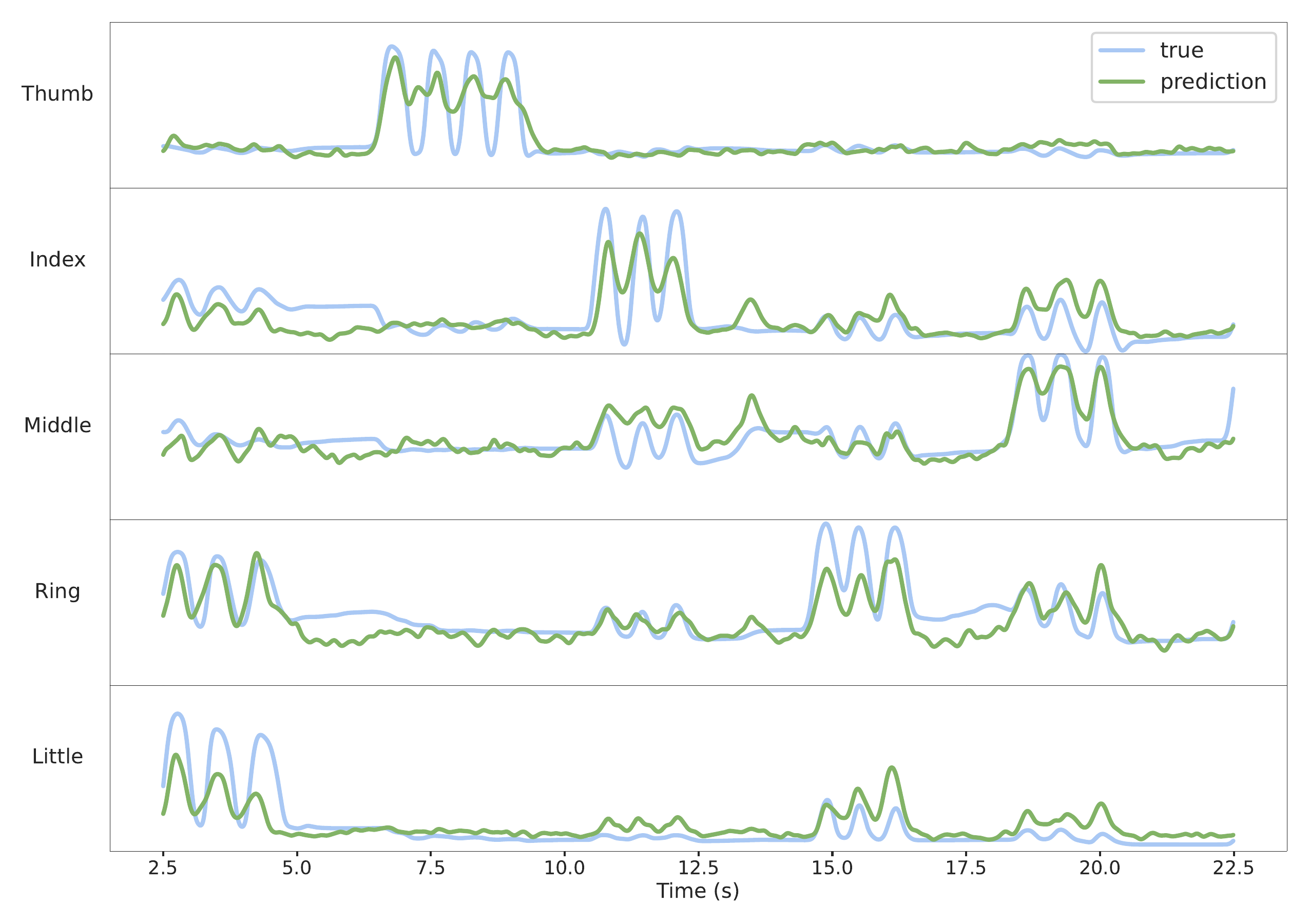}
    \caption{An example of true and decoded finger trajectories time series. The data segment for visualization was taken from the validation set}
    \label{fig:time series}
\end{figure*}


Our approach shows great promise in developing high-precision decoders of neural signals related to movements for use in the brain-computer interface. The proposed model significantly outperforms the winner of the BCI competition IV with a mean correlation value of approximately $0.67$. It also improves greatly the current state-of-the-art results of other research groups solving the finger trajectory decoding task (Table \ref{table:bcicomp_comparison}). Moreover, FingerFlex successfully decoded finger trajectories using the other dataset without any additional hyperparameter search (Table \ref{table:stanford_performance})

The method developed in this study stands on several important ideas and improvements. First of all, the time-frequency feature selection choice was crucial to capture fine changes in different frequency components of the brain signals responsible for finger movements. Secondly, the hybrid encoder-decoder architecture with skip connections was employed for the first time for the ECoG decoding task. In addition, the introduction of a new loss function to a conventional MSE further improved regression results. Lastly, the physiologically relevant hyperparameters, such as time delay between finger trajectory and ECoG time series and frequency range for wavelet convolution alongside with conventional hyperparameters, such as learning rate and model depth, are stable across participants and datasets. These hyperparameters can be fixed once providing sufficient performance for the task.  

There are also several advantages inherent in the suggested approach. Firstly, our method does not require training separate decoders for resting and moving states, as that was done in \citet{elgharabawy2016} and \citet{flamary2012}. It captures both changes in finger coordinate during movement periods and during resting periods without switching and adding any additional information. Secondly, the model is relatively lightweight, having only about $600 000$ parameters, which potentially allows to use it in real-time inference applications. Furthermore, the method does not rely on ECoG data, which follows the movement in time, i.e. it makes predictions based only on a current time window, not taking into account future and preceding ones. This feature further facilitates real-time deployment. Lastly, the length of the time window can be arbitrary due to the nature of convolution operation, thus providing further flexibility for other decoding solutions. 

\subsection{Limitations and potential improvements}

The unequal electrode placement between subjects and uncertainty about electrode positioning relative to the brain limit the possible generalization approaches. However, the proposed method is highly suitable for individualized training. One of the strongest arguments for that is that only 6.5 minutes of recording is sufficient to achieve a significant performance on decoding fine movements. 

One of the further research directions is implementing transfer learning across participants \citep{wan2021}. If electrode positions and subjects' magnetic-resonance imaging scans are available, it becomes possible to use source reconstruction approaches \citep{hamalainen1993, chen2002}, i.e. solving the "inverse problem". This allows making a transition from individualized brain surface spatial patterns (electrode space) to standardized cortical activations (source space). This is essential for developing cross-subjects solutions and training a network on the larger amounts of data gathered in different experimental settings and across participants. 
Another possible way of solving the lack of correspondence of electrode location between subjects is to use separate feature extractors for each subject inferring generalized features suitable for transfer learning like was done in the work of \citet{peterson2021}. 

Besides transfer learning, a possible advancement could be adding interpretable blocks to enhance understanding of the data. An important step in that direction was made by \citet{petrosyan2021} and \citet{sturm2016}, and while adding interpretable blocks might worsen model performance, research in this direction could provide useful insights on motor encoding in the human brain and what features of ECoG signals really matter for the movement prediction.

Finally, an important step would be to omit any manual feature extraction of the time series data allowing the neural network to learn to extract relevant features for the task itself. Time-frequency methods have their limitations and their use imply increasing a number hyperparameters, which is not always favorable. Temporal convolutions are by far the most relevant way to implement automatic feature extraction and this approach also works well, as was shown by \citet{li2021a}, \citet{li2021} and \citet{peterson2021}. The possible downside of this approach would be the increased demand for lengthy data due to the increased number of parameters in the network, and gathering more data is often impossible in clinical settings of neurophysiological research.  

\section{Conclusion}

The present study developed a novel deep learning approach for decoding finger flexions based on oscillatory activity on the ECoG channels. With the use of convolutional encoder-decoder architecture with skip-connections, the performance on the trajectory regression task was significantly improved compared to the existing solutions using the BCI competition IV dataset 4. The proposed model can potentially be used for real-time inference of finger flexions. This research shows promise for developing fast, functional, and precise neuroprosthetic devices. 

\bibliography{refs} 

\end{document}